\documentstyle[12pt,epsf]{article}

\begin{document}

\baselineskip 6mm
\renewcommand{\thefootnote}{\fnsymbol{footnote}}

%------------ Hyun Seok's macro's, etc  -----------

\newcommand{\nc}{\newcommand}
\newcommand{\rnc}{\renewcommand}

%\headheight=0truein
%\headsep=0truein
%\topmargin=0truein
%\oddsidemargin=0truein
%\evensidemargin=0truein
%\textheight=9truein
%\textwidth=6.5truein

\rnc{\baselinestretch}{1.24}    % 1.5 spacing btwn text lines
\setlength{\jot}{6pt}       % spacing btwn the rows of an eqnarray
\rnc{\arraystretch}{1.24}       % spacing btwn the rows of a non-eqn array

%%%%%%%%%%%%%%%%%%%%%% Equation Numbering %%%%%%%%%%%%%%%%%%%%%%%
\makeatletter
\rnc{\theequation}{\thesection.\arabic{equation}}
\@addtoreset{equation}{section}
\makeatother

%%%%%%%%%%%%%%%%%%%%%%%%%%%%%%%%%%%%%%%%%%%%%%%%%%%%%%%%%%%%%%%%%
%                               %
%       NEW COMMANDS AND MACROS             %
%                               %
%%%%%%%%%%%%%%%%%%%%%%%%%%%%%%%%%%%%%%%%%%%%%%%%%%%%%%%%%%%%%%%%%

%%%%% Simplify some frequently used LaTeX commands %%%%%

\nc{\be}{\begin{equation}}
\nc{\ee}{\end{equation}}
\nc{\bea}{\begin{eqnarray}}
\nc{\eea}{\end{eqnarray}}
\nc{\xx}{\nonumber\\}
\nc{\ct}{\cite}
\nc{\la}{\label}
\nc{\eq}[1]{(\ref{#1})}
\nc{\newcaption}[1]{\centerline{\parbox{6in}{\caption{#1}}}}
\nc{\fig}[3]{

\begin{figure}
\centerline{\epsfxsize=#1\epsfbox{#2.eps}}
\newcaption{#3. \label{#2}}
\end{figure}
}

%%% Caligraphic letters %%%%

\def\CA{{\cal A}}
\def\CC{{\cal C}}
\def\CD{{\cal D}}
\def\CE{{\cal E}}
\def\CF{{\cal F}}
\def\CG{{\cal G}}
\def\CH{{\cal H}}
\def\CK{{\cal K}}
\def\CL{{\cal L}}
\def\CM{{\cal M}}
\def\CN{{\cal N}}
\def\CO{{\cal O}}
\def\CP{{\cal P}}
\def\CS{{\cal S}}
\def\CU{{\cal U}}
\def\CW{{\cal W}}
\def\CY{{\cal Y}}

%%% Double line letters %%%

\def\IR{{\hbox{{\rm I}\kern-.2em\hbox{\rm R}}}}
\def\IB{{\hbox{{\rm I}\kern-.2em\hbox{\rm B}}}}
\def\IN{{\hbox{{\rm I}\kern-.2em\hbox{\rm N}}}}
\def\IC{\,\,{\hbox{{\rm I}\kern-.59em\hbox{\bf C}}}}
\def\IZ{{\hbox{{\rm Z}\kern-.4em\hbox{\rm Z}}}}
\def\IP{{\hbox{{\rm I}\kern-.2em\hbox{\rm P}}}}
\def\IH{{\hbox{{\rm I}\kern-.4em\hbox{\rm H}}}}
\def\ID{{\hbox{{\rm I}\kern-.2em\hbox{\rm D}}}}

%%% Greek letters %%%

\def\a{\alpha}
\def\b{\beta}
\def\ga{\gamma}
\def\d{\delta}
\def\ep{\epsilon}
\def\ph{\phi}
\def\k{\kappa}
\def\l{\lambda}
\def\m{\mu}
\def\n{\nu}
\def\th{\theta}
\def\rh{\rho}
\def\s{\sigma}
\def\t{\tau}
\def\w{\omega}
\def\G{\Gamma}

%%%%% Mathematical Symbols

\def\half{\frac{1}{2}}
\def\dint#1#2{\int\limits_{#1}^{#2}}
\def\goto{\rightarrow}
\def\para{\parallel}
\def\brac#1{\langle #1 \rangle}
\def\grad{\nabla}
\def\curl{\nabla\times}
\def\div{\nabla\cdot}
\def\p{\partial}
\def\e{\epsilon_0}

%%%%% Roman pont in math

\def\Tr{{\rm Tr}\,}
\def\det{{\rm det}}

%%%%% Special Letters

\def\vare{\varepsilon}
\def\bz{\bar{z}}
\def\bw{\bar{w}}

%%%%% For this paper only

\def\do{{\bf R}^2_{NC}}
\def\re{{\bf R}^2_C}
\def\mi{{\bf R}^4_{NC}}
\def\c{{\bf C}}
\def\z{{\bf Z}}

\begin{titlepage}

%---------------- preprint number ---------------
\hfill\parbox{4cm} {{\tt hep-th/0109121}}

\vspace{25mm}

\begin{center}
%------------------------ title ------------------------

{\Large \bf Noncommutative Instantons on
$\do \times \re$}\\

\vspace{15mm}
%---------------- authors and addresses ----------------
Keun-Young Kim$^a$\footnote{kykim@physics4.sogang.ac.kr},
Bum-Hoon Lee$^a$\footnote{bhl@ccs.sogang.ac.kr} and Hyun Seok
Yang$^b$\footnote{hsyang@phys.ntu.edu.tw}
\\[10mm]
$^a${\sl Department of Physics, Sogang University \\
Seoul 121-742, Korea}\\
$^b${\sl Department of Physics, National Taiwan University \\
Taipei 106, Taiwan, R.O.C.}

\end{center}

\thispagestyle{empty}

\vskip 20mm

%----------------------- abstract ----------------------
\centerline{\bf ABSTRACT}

\vskip 4mm

\noindent We study $U(1)$ and $U(2)$ noncommutative instantons on
$\do \times \re$ based on the ADHM construction. It is shown that
a mild singularity in the instanton solutions for both self-dual
and anti-self-dual gauge fields always disappears in gauge
invariant quantities and thus physically regular solutions can be
constructed even though any projected states are not involved in
the ADHM construction. Furthermore the instanton number is also an
integer.

\vspace{2cm}
\end{titlepage}

%-------------------------------------------------------
\baselineskip 7mm
\renewcommand{\thefootnote}{\arabic{footnote}}
\setcounter{footnote}{0}

%%%%%%%%%%%%%%%%%%%%%%%%%%%%%%%%%%%%%%%%%%%%%%%%%%%%%%%%%%%%%%%%%%%%%%
\section{Introduction}
%%%%%%%%%%%%%%%%%%%%%%%%%%%%%%%%%%%%%%%%%%%%%%%%%%%%%%%%%%%%%%%%%%%%%%
A noncommutative space is obtained by quantizing a given space
with its symplectic structure, treating it as a phase space. Also
field theories can be formulated on a noncommutative space.
Noncommutative field theory means that fields are defined as
functions over noncommutative spaces. At the algebraic level, the
fields become operators acting on a Hilbert space as a
representation space of the noncommutative space. Since the
noncommutative space resembles a quantized phase space, the idea
of localization in ordinary field theory is lost. The notion of a
point is replaced by that of a state in representation space.

Instanton solutions in noncommutative Yang-Mills theory can also
be studied by Atiyah-Drinfeld-Hitchin-Manin (ADHM) equation
\ct{adhm} slightly modified by the noncommutativity. Recently much
progress has been made in this direction
\ct{ns,mb,naka,sw,ly,bn,kf1,kf2,kf3,kko,ho,kly,lty,nek,as,corr,ckt,ham,kuro}.
The remarkable fact is that the deformation of the ADHM equation
depends on the self-duality of the noncommutativity \ct{sw}.
Anti-self-dual instantons on self-dual noncommutative $\mi$ are
described by a deformed ADHM equation adding a Fayet-Iliopoulos
term to the usual ADHM equation and the singularity of instanton
moduli space is resolved \ct{ns,mb}. However, self-dual instantons
on self-dual $\mi$ are described by an undeformed ADHM equation
and the singularity of instanton moduli space still remains
\ct{kf3,ckt}. This property is closely related to the BPS property
of D0-D4 system \ct{sw}. The latter system is supersymmetric and
BPS whereas the former is not BPS.

In this report, we study $U(1)$ and $U(2)$ noncommutative
instantons on $\do \times \re$ based on the ADHM construction
where ${\bf R}^2_{NC}$ is the noncommutative space but ${\bf
R}^2_{C}$ is the commutative space. It was already shown in
\ct{kly} that the completeness relation in the ADHM construction
is generally satisfied for $\do \times \re$ as well as $\mi$.
Unlike $\mi$, the ADHM equation for the noncommuative space $\do
\times \re$ is always deformed for self-dual and anti-self-dual
gauge fields since both systems are not BPS states any more. This
implies that the small instanton singularity of moduli space can
be resolved for this case too. Actually, even though the instanton
solutions for both self-dual and anti-self-dual gauge fields
contain a mild singularity, i.e. a measure zero singularity, it
always disappears in gauge invariant quantities and thus
physically regular solutions can be constructed even though any
projected states are not involved in the ADHM construction.
Furthermore the instanton number is always an integer as it
should be. Our result is different from \ct{ckt} by Chu, et al.
claiming that there is no nonsingular $U(N)$ instanton on ${\bf
R}^2_{NC} \times {\bf R}^2_{C}$ due to the breakdown of the
completeness relation.

The paper is organized as follows. In next section we briefly
review the ADHM construction of noncommutative instantons on $\do
\times \re$. In section 3 we explicitly calculate the self-dual
and anti-self-dual field strengths for single $U(1)$ and $U(2)$
instantons. We show that physically non-singular solutions can be
constructed and they correctly give integer instanton numbers. In
section 4 we discuss the results obtained and address some issues.

%%%%%%%%%%%%%%%%%%%%%%%%%%%%%%%%%%%%%%%%%%%%%%%%%%%%%%%%%%%%%%%%%%%%%%
\section{ADHM Equations on $\do \times \re$}
%%%%%%%%%%%%%%%%%%%%%%%%%%%%%%%%%%%%%%%%%%%%%%%%%%%%%%%%%%%%%%%%%%%%%%

Let's briefly review the ADHM construction on ${\bf R}^2_{NC}
\times {\bf R}^2_{C}$ where ${\bf R}^2_{NC}$ is the
noncommutative space but ${\bf R}^2_{C}$ is the commutative
space. This space is represented by the algebra generated by
$x^\mu$ obeying the commutation relation:
\begin{equation} \la{nc2c2}
[x^\mu, x^\nu]=i\theta^{\mu\nu},
\end{equation}
where $\mu,\,\nu=1,2,3,4$ and the matrix $\theta^{\mu\nu}$ is of
rank-two. We set here $\theta^{12}={\zeta \over 2}=-\theta^{21}$
and $\theta^{34}=0$. In terms of complex coordinates
$$ z_1 = x^2
+ i x^1, \quad z_2 = x^4 + i x^3,$$ the commutation relation
\eq{nc2c2} reduces to
\begin{equation}
\label{dore}
 [\bar{z}_1, z_1]=\zeta, \;\;\;
 [\bar{z}_2, z_2] =0,
\end{equation}
which generates an operator algebra denoted as $\CA$. The
commuatation algebra on $\bar{z}_1,\, z_1$ is that of a simple
harmonic oscillator and so one may use the Hilbert space $\CH=
\sum_{n \in {\bf Z}_{\geq 0}} {\bf C} | n>$ as a representation
of this algebra, where $\bar{z}_1,\, z_1$ are represented as the
annihilation and the creation operators:
\begin{equation} \label{sho}
\sqrt{\frac{1}{\zeta} } \bar{z}_1 | n> = \sqrt{n} | n - 1>, \quad
\sqrt{\frac{1}{\zeta} } z_1 | n> = \sqrt{n+1 } | n +1>.
\end{equation}
Thus the integration on $\do \times \re$ for an operator ${\cal
O}(x)$ in $\CA$ can be replaced by
\begin{equation}
\la{trace} \int d^4x {\cal O}(x) \rightarrow \zeta\pi \sum_{n \in
{\bf Z}_{\geq 0}} \int d^2x \langle n|{\cal O}(x)|n \rangle,
\end{equation}
where $d^2x=dx^3 dx^4$.

ADHM construction describes an algebraic way for finding
(anti-)self-dual configurations of the gauge field in terms of
some quadratic matrix equations on four manifolds \ct{adhm}. The
ADHM construction can be generalized to the space $\do \times \re$
under consideration \cite{ckt}. In order to describe $k$
instantons with gauge group
$U(N)$, one starts with the following data:\\
1. A pair of complex hermitian vector spaces
$V={\bf C}^k,\;W={\bf C}^N$.\\
2. The operators $B_1, B_2 \in Hom(V,V),\;I \in Hom(W,V)$ and $J
\in Hom(V,W)$ satisfying the equations \begin{eqnarray} \la{ADHMr}
&&\mu_r = [B_1 , B_1 ^\dagger ] +[B_2 , B_2 ^\dagger ] +
II^\dagger -
J^\dagger J=\zeta,\\
\la{ADHMc} &&\mu_c = [B_1 , B_2 ] + IJ=0.
\end{eqnarray}
3. Define a Dirac operator $D^\dagger : V \oplus V \oplus W
\rightarrow V \oplus V $ by
\begin{equation}
D^\dagger = \pmatrix { \tau_z \cr \sigma^\dagger_z }
\end{equation}
where
\begin{equation}\la{ASD}
\tau_z = \pmatrix { B_2 - z_2 & B_1 - z_1 & I }, \quad \sigma_z =
\pmatrix {-B_1 + z_1 \cr B_2 - z_2 \cr J }
\end{equation}
for anti-self-dual instantons and
\begin{equation}\la{SD}
\tau_z = \pmatrix { B_2 - {\bar z}_2 & B_1 + z_1 & I }, \quad
\sigma_z = \pmatrix {-B_1 - z_1 \cr B_2 - {\bar z}_2 \cr J }
\end{equation}
for self-dual instantons.

The origin of the ADHM equations \eq{ADHMr} and \eq{ADHMc} is the
so-called factorization condition:
\begin{equation} \la{factorization}
\tau_z\tau^\dagger_z=\sigma^\dagger_z\sigma_z, \quad
\tau_z\sigma_z=0.
\end{equation}
Note that, unlike $\mi$, the ADHM equation \eq{ADHMr} for $\do
\times \re$ is always deformed for self-dual and anti-self-dual
instantons. According to the ADHM construction, one can get the
gauge field (instanton solution) by the formula
\begin{equation}
\la{A} A_\mu = \psi^\dagger \partial_\mu \psi,
\end{equation}
where $\psi : W \rightarrow V \oplus V \oplus W $ is $N$
zero-modes of $D^\dagger$, i.e.,
\begin{equation}
\label{zmpsi} D^\dagger \psi = 0.
\end{equation}
For given ADHM data and the zero mode condition \eq{zmpsi}, the
following completeness relation has to be satisfied to construct
(anti-)self-dual instantons from the gauge field \eq{A}
\begin{equation}\label{CR}
  D{1\over D^\dagger D}D^\dagger + \psi \psi^\dagger = 1.
\end{equation}
It was shown in \ct{kly} that this relation is always satisfied
even for noncommutative spaces.

The space $\do \times \re$ doesn't have any isolated singularity
due to the factor $\do$. However in this case it is a measure
zero singularity, so it doesn't cause any physical trouble
although we don't project out it. This property presents a
striking contrast to $\mi$ \ct{kf1,kf2,kly,ckt} since this space
is two-dimensional discrete lattice, so singularities are always
separable. Actually, it will be shown that, for the space $\do
\times \re$, the singularity in the instanton solutions always
disappears in the gauge invariant quantities, e.g. $\Tr_{\CH}
F^n$ where $\Tr_{\CH}$ is the integration over $\do$, possibly
including the group trace too. So in our ADHM construction we
will not project out any state in $\CH$ \footnote{If one insists
on projecting out some states in $\CH$, e.g. $|0\rangle$, then the
whole $\re$ plane at $|0\rangle$ is necessarily projected out. A
serious trouble here is that one cannot have a projection
projecting out only offending states, e.g. $|0\rangle$ at
$z_2=0$. This too excessive projection causes the breaking of the
completeness relation \eq{CR} as shown in \ct{ckt}. We thank the
authors of \ct{ckt} for this discussion.} and thus the zero-modes
\eq{zmpsi} are normalized in usual way
\begin{equation}
\la{norm=1}
\psi^\dagger \psi=1.
\end{equation}
With the above relations, the anti-self-dual field strength
$F_{ASD}$ can be calculated by the following formula
\begin{eqnarray}\la{asdF}
F_{ASD}&=&\psi^\dagger\left(d\tau_z^\dagger{1\over\triangle_z}d\tau_z
+d\sigma_z{1\over\triangle_z}d\sigma_z^\dagger\right)\psi
\nonumber \\
\label{ASDF} &=& \psi^\dagger
\pmatrix{dz_1{1\over\triangle_z}d\bar{z}_1 -
dz_2{1\over\triangle_z}d{\bar z}_2 & - 2
dz_1{1\over\triangle_z}d\bar{z}_2 &  0  \cr
  -2 dz_2 {1\over\triangle_z}d{\bar z}_1 &
  -dz_1{1\over\triangle_z}d{\bar z}_1
+ dz_2{1\over\triangle_z}d\bar{z}_2 &  0  \cr
                           0 & 0 & 0 }    \psi,
\end{eqnarray}
where $\triangle_z=\tau_z\tau^\dagger_z=\sigma^\dagger_z\sigma_z$
has no zero-modes so it is invertible. Similarly, the self-dual
field strength $F_{SD}$ can be calculated by
\begin{eqnarray}\la{sdF}
F_{SD}&=&\psi^\dagger\left(d\tau_z^\dagger{1\over\triangle_z}d\tau_z
+d\sigma_z{1\over\triangle_z}d\sigma_z^\dagger\right)\psi
\nonumber \\
\label{SDF} &=& \psi^\dagger
\pmatrix{dz_1{1\over\triangle_z}d\bar{z}_1 +
dz_2{1\over\triangle_z}d{\bar z}_2 & 2 dz_1{1\over\triangle_z}dz_2
& 0 \cr -2 d\bar{z}_1{1\over\triangle_z}d\bar{z}_2 &
  - dz_1{1\over\triangle_z} d\bar{z}_1
- dz_2{1\over\triangle_z}d{\bar z}_2 &  0  \cr
                           0 & 0 & 0 }    \psi.
\end{eqnarray}

%%%%%%%%%%%%%%%%%%%%%%%%%%%%%%%%%%%%%%%%%%%%%%%%%%%%%%%%%%%%%%%%%%%%%%
\section{Instanton Solutions on $\do \times \re$}
%%%%%%%%%%%%%%%%%%%%%%%%%%%%%%%%%%%%%%%%%%%%%%%%%%%%%%%%%%%%%%%%%%%%%%

In this section we will solve the ADHM equation \eq{zmpsi} for
single $U(1)$ and $U(2)$ instantons and calculate both the
anti-self-dual field strength \eq{asdF} and the self-dual field
strength \eq{sdF}. Also we will numerically calculate the
topological charge for the solutions to show it is always an
integer. It naturally turns out that, even though any state in
$\CH$ is not projected out, a mild singularity in the solution
doesn't cause any physical trouble and they can define physically
regular solutions.

\subsection{Anti-self-dual $U(1)$ Instanton}

In this case the ADHM equation \eq{zmpsi} can be solved in the
exactly same way as the $\mi$ case, only keeping in mind the
algebra \eq{dore}. The solution $\psi = \psi_1 \oplus \psi_2
\oplus \xi$ in $V \oplus V \oplus W$ has the same form as the
anti-self-dual instanton on self-dual $\mi$ \ct{kly}:
\begin{equation}\la{asdu1}
\psi_1 = \bar{z}_2\sqrt{\zeta \over \delta\Delta}, \quad \psi_2 =
\bar{z}_1\sqrt{\zeta \over \delta\Delta}, \quad \xi=\sqrt{\delta
\over \Delta},
\end{equation}
where $\delta = z_1\bar{z}_1+ z_2\bar{z}_2$ and $\Delta =\delta
+\zeta$.

It is straightforward to calculate the anti-self-dual field
strength $F_{ASD}$ for the solution \eq{asdu1} from \eq{asdF}:
\begin{eqnarray}
\label{asdf1} F_{ASD} &=&{\zeta\over \delta^2 \Delta^2}(z_1\bar
z_1 \Delta-z_2\bar z_2 \delta)(dz_2 d\bar{z}_2 - dz_1 d\bar{z}_1)
\xx &+& {2\zeta\over \Delta^2\sqrt{\delta(\Delta + \zeta)}}
z_2\bar z_1 d\bar{z}_2 dz_1 +{2\zeta\over
\delta^2\sqrt{\Delta(\delta - \zeta)}} z_1\bar z_2 d\bar{z}_1
dz_2.
\end{eqnarray}
One can see that the above field strength contains a (mild)
singularity of the type $z_2/|z_2|$ in the second term at the
state $|1\rangle$ and $z_2=0$ and in the third term at the state
$|0\rangle$ and $z_2=0$. These singularities are placed at single
point, the origin, of $\re$ only at $|0\rangle$ or $|1\rangle$.
So these are measure zero singularities, which is very similar
situation to the usual singular gauge $SU(2)$ instantons. There
is no huge plane singularity claimed in \ct{ckt}.

Note that the field strength $F_{ASD}$ in noncommutative gauge
theory is not a gauge invariant quantity. Rather the gauge
invariant quantity is $\Tr_{\CH} F_{ASD}$ which is definitely
singularity-free. Also the instanton density defined below is
singularity-free. See the Fig. 1-a. Thus one can see that the
singularity in \eq{asdf1} doesn't cause any physical trouble and
the physical quantities such as $\Tr_{\CH} F^n$ are well-defined
although there is no projected state in $\CH$.

Finally the topological charge can be easily calculated by using
the prescription \eq{trace}
\begin{eqnarray}\la{asdQ1}
Q &=&- \sum_{n=0}^\infty \int_0^\infty d\gamma Q_n(\gamma)\\
&=& -
\sum_{n=0}^\infty \int_0^\infty d\gamma
\Biggl[{\Bigl(n(n+1+\gamma)-(n+\gamma)\gamma\Bigr)^2\over
(n+\gamma)^4(n+1+\gamma)^4} + {4\gamma(n+1)\over
(n+\gamma)(n+1+\gamma)^4(n+2+\gamma)}\Biggr]\xx &=&-1\nonumber,
\end{eqnarray}
where $n=z_1\bar z_1/\zeta$ and $\gamma=(x_3^2+x^2_4)/\zeta$. If
one projects out the state $|0\rangle$, one could not get $-1$
for $Q$.

\subsection{Anti-self-dual $U(2)$ Instanton}

For this case also the ADHM equation \eq{zmpsi} can be solved in
the exactly same way and the solution has the same form as the
$\mi$ case \ct{kly}:
\begin{equation}\la{asdu2}
 \psi = \pmatrix{\psi^{(1)} & \psi^{(2)}} =
\pmatrix{\bar z_2 f & -z_1 g\cr
  \bar z_1 f & z_2 g \cr \xi_1 & 0 \cr 0 & \xi_2 },
\end{equation}
where
\begin{equation}\label{fg}
  f=\sqrt{\rho^2+\zeta \over \delta(\Delta+\rho^2)},\;\;\;
  g=\sqrt{\rho^2 \over \Delta(\Delta+\rho^2)},\;\;\;
  \xi_1=\sqrt{\delta \over \Delta+\rho^2},\;\;\;
  \xi_2=\sqrt{\Delta \over \Delta+\rho^2}\,.
\end{equation}
When the instanton size vanishes, that is $\rho = 0$, then $g= 0$,
and, from \eq{asdF}, one can see that $\psi^{(2)}$ does not
contribute to the field strength. Therefore the structure of the
$U(2)$ instanton at $\rho = 0$ is completely determined by the
minimal zero-mode $\psi^{(1)}$ in the $U(1)$ subgroup. This
property is exactly same as the $\mi$ case \ct{kf1,kly}.

The field strength $F_{ASD}$ can be obtained from \eq{asdF} with
the solution \eq{asdu2}:
\begin{eqnarray}\la{asdf2}
F_{ASD}&=&(d\bar{z}_2 \wedge dz_2 -d\bar{z}_1 \wedge dz_1)
\pmatrix{a_{11} & a_{12} \cr
 a_{12}^\dag & a_{22}} \nonumber \\
&+&d\bar{z}_1 \wedge dz_2 \pmatrix{b_{11} & b_{12} \cr
 b_{21} & b_{22}} - dz_1 \wedge d\bar{z}_2
 \pmatrix{b_{11}^\dag & b_{21}^\dag \cr
 b_{12}^\dag & b_{22}^\dag},
\end{eqnarray}
where
\begin{eqnarray}
&&a_{11}={\rho^2+\zeta \over
\delta(\delta+\rho^2)(\Delta+\rho^2)^2}
\Bigl((\delta+\rho^2)z_2\bar z_2 - (\Delta+\rho^2)z_1\bar
z_1\Bigr), \xx && a_{12}=-{1\over
\delta}\sqrt{\rho^2(\rho^2+\zeta) \over
(\delta+\rho^2)(\Delta+\rho^2)} \Bigl({1 \over \delta+\rho^2} +
{1 \over \Delta+\rho^2} \Bigr)z_1z_2,\xx && a_{22}={\rho^2 \over
\Delta(\Delta+\rho^2)^2(\Delta+\rho^2+\zeta)}
\Bigl((\Delta+\rho^2)(z_1\bar z_1+\zeta) -
(\Delta+\rho^2+\zeta)z_2\bar z_2\Bigr),\xx &&
b_{11}={2(\rho^2+\zeta) \over \delta+\rho^2} \sqrt{1\over
(\delta-\zeta)\delta(\delta+\rho^2)(\Delta+\rho^2)}\, z_1\bar
z_2,\\ && b_{12}=-{2\over \delta+\rho^2}
\sqrt{\rho^2(\rho^2+\zeta) \over
(\delta-\zeta)\delta(\delta+\rho^2-\zeta)(\Delta+\rho^2)}\, z_1
z_1,\xx && b_{21}={2\over (\Delta+\rho^2)^2}
\sqrt{\rho^2(\rho^2+\zeta) \over \delta \Delta}\, \bar z_2 \bar
z_2,\xx && b_{22}=-{2 \rho^2 \over \Delta+\rho^2} \sqrt{1 \over
\delta(\delta+\rho^2)\Delta(\Delta+\rho^2)}\, z_1 \bar
z_2.\nonumber
\end{eqnarray}
It can be confirmed again to recover the ordinary $SU(2)$
instanton solution in the $\zeta = 0$ limit and the $U(1)$
solution \eq{asdf1} for the limit $\rho =0$ where only $a_{11}$
and $b_{11}$ terms in \eq{asdf2} survive. It is a pleasant
property that the solution shows smooth behaviors with respect to
$\rho$ and $\zeta$ (except only $\rho=\zeta=0$).

One can explicitly check that the field strength \eq{asdf2} has
the exactly same kind of singularity appeared in \eq{asdf1} and
it appears only in $b_{11}$ and in $b_{11}^\dag$ which is just
$U(1)$ part. Thus this singularity doesn't cause any physical
trouble either for the exactly same reason in section 3.1.

After a little but straightforward algebra, one can determine the
instanton charge density $Q_n(\gamma)$ and calculate the
topological charge of the solution \eq{asdf2}:
\begin{equation}\la{asdQ2}
Q = - \sum_{n=0}^\infty \int_0^\infty d\gamma Q_n(\gamma),
\end{equation}
where
\begin{eqnarray}\la{densityQ}
Q_n(\gamma)&=& {1\over (n+1+\gamma+a^2)^4}\biggl(
{(1+a^2)^2\gamma^2 \over (n+\gamma)^2} + {a^4 \gamma^2 \over
(n+1+\gamma)^2} + {4a^2(1+a^2) \gamma^2 \over
(n+\gamma)(n+1+\gamma)}\xx &-& {2(1+a^2)^2 n \gamma
(n+1+\gamma+a^2) \over (n+\gamma)^2(n+\gamma+a^2)}+ {4a^4 n \gamma
(n+1+\gamma+a^2) \over (n+\gamma)(n+1+\gamma)(n+\gamma+a^2)}\xx
&+&{(1+a^2)^2 n^2 (n+1+\gamma+a^2)^2 \over
(n+\gamma)^2(n+\gamma+a^2)^2}+ {4(1+a^2)^2 n \gamma
(n+1+\gamma+a^2)^3 \over
(n-1+\gamma)(n+\gamma)(n+\gamma+a^2)^3}\xx &+& {4a^2(1+a^2)(n-1) n
(n+1+\gamma+a^2)^3 \over
(n-1+\gamma)(n+\gamma)(n-1+\gamma+a^2)(n+\gamma+a^2)^2} \xx
&+&{a^4 (n+1)^2 (n+1+\gamma+a^2)^2 \over
(n+1+\gamma)^2(n+2+\gamma+a^2)^2}- {2a^4 (n+1) \gamma
(n+1+\gamma+a^2) \over (n+1+\gamma)^2(n+2+\gamma+a^2)}\xx &+&
{2a^2(1+a^2) n \gamma (n+1+\gamma+a^2)(2n+1+2\gamma+2a^2)^2 \over
(n+\gamma)^2(n+\gamma+a^2)^3}\biggr)
\end{eqnarray}
with $a=\rho/\sqrt\zeta$. The charge density $Q_n(\gamma)$ is
smooth function with respect to $\gamma$ for all $n$. See the
Fig. 1-b, c, d. We performed the integral first and then the summation in
\eq{asdQ2} numerically using Mathematica and the result is
summarized below (where we indicate the summation range for each
case).
\begin{equation}\la{table1}
\begin{tabular}{|c|c|c|c|}
  \hline
    & $a=0.1$ ($0\leq n \leq 10^2$) & $a=1$ ($0\leq n \leq 10^2$)
    & $a=10$  ($0\leq n \leq 10^3$) \\
    \hline
  $Q$ & $-0.998541$ & $-0.999791$ & $-0.991667$ \\ \hline
\end{tabular}
\end{equation}
We further checked that $Q(100) \equiv - \sum_{n=0}^{100}
\int_0^\infty d\gamma Q_n(\gamma)=\sum_{k=0} q_k a^{2k}$, and we
obtained $q_0=-0.998542,\;q_1=9.82 \times 10^{-5},\; q_2=-9.63
\times 10^{-5}$, etc.

Noting that the topological charge density $Q_n(\gamma)$ in
\eq{densityQ} is rapidly convergent series with respect to $n$
after the $\gamma$-integration, the above numerical results lead
us to the conclusion very confidently that the topological charge
of the anti-self-dual $U(2)$ instanton is also an integer and
independent of the modulus $\rho$.

\subsection{Self-dual $U(1)$ Instanton}

Now we will solve the ADHM equation \eq{zmpsi} with the self-dual
ADHM data \eq{SD}. The solution can be found very easily:
\begin{equation}\la{sdu1}
\psi_1 = z_2\sqrt{\zeta \over \delta\Delta}, \quad \psi_2 =
-\bar{z}_1\sqrt{\zeta \over \delta\Delta}, \quad \xi=\sqrt{\delta
\over \Delta}.
\end{equation}
Also one can easily calculate the self-dual field strength
$F_{SD}$ for the solution \eq{sdu1} from \eq{sdF}:
\begin{eqnarray}
\label{sdf1} F_{SD} &=&-{\zeta\over \delta^2 \Delta^2}(z_1\bar
z_1 \Delta-z_2\bar z_2 \delta)(dz_1 d\bar{z}_1 + dz_2 d\bar{z}_2)
\xx &-& {2\zeta\over \Delta^2\sqrt{\delta(\Delta + \zeta)}} \bar
z_1\bar z_2 dz_1 dz_2 +{2\zeta\over \delta^2\sqrt{\Delta(\delta -
\zeta)}} z_1 z_2 d\bar{z}_1 d\bar z_2.
\end{eqnarray}
It can be checked explicitly that the self-dual field strength is
also well-defined for all states in $\CH$ and on $\re$ except the
mild singularities of the second and the third terms. But, for the
same reason as the previous cases, this singularity is never
harmful and we can well define singularity-free physical
quantities such as $\Tr_{\CH} F^n$ from $F_{SD}$ in \eq{sdf1}.

The topological charge for the solution \eq{sdf1} has the same
expression as \eq{asdQ1} except the sign which is now $+$, so we
get $Q=1$.

\subsection{Self-dual $U(2)$ Instanton}

The self-dual $U(2)$ instantons can be obtained by solving the
ADHM equation \eq{zmpsi} with the data \eq{SD}:
\begin{equation}\la{sdu2}
 \psi = \pmatrix{\psi^{(1)} & \psi^{(2)}} =
\pmatrix{z_2 f & z_1 g\cr
  -\bar z_1 f & \bar z_2 g \cr \xi_1 & 0 \cr 0 & \xi_2 }
\end{equation}
with the notation \eq{fg}. For the same reason as section 3.2 the
structure of the self-dual instanton at $\rho = 0$ is also
completely determined by the minimal zero-mode $\psi^{(1)}$ in
the $U(1)$ subgroup.

The field strength $F_{SD}$ for the solution \eq{sdu2} can be
easily obtained from \eq{sdF}:
\begin{eqnarray}\la{sdf2}
F_{SD}&=&(dz_1 \wedge d\bar z_1 +dz_2 \wedge d\bar z_2)
\pmatrix{c_{11} & c_{12} \cr
 c_{12}^\dag & c_{22}} \nonumber \\
&+&dz_1 \wedge dz_2 \pmatrix{d_{11} & d_{12} \cr
 d_{21} & d_{22}} + d\bar z_1 \wedge d\bar{z}_2
 \pmatrix{d_{11}^\dag & d_{21}^\dag \cr
 d_{12}^\dag & d_{22}^\dag},
\end{eqnarray}
where
\begin{eqnarray}
&&c_{11}={\rho^2+\zeta \over
\delta(\delta+\rho^2)(\Delta+\rho^2)^2}
\Bigl((\delta+\rho^2)z_2\bar z_2 - (\Delta+\rho^2)z_1\bar
z_1\Bigr), \xx && c_{12}={1\over
\delta}\sqrt{\rho^2(\rho^2+\zeta) \over
(\delta+\rho^2)(\Delta+\rho^2)} \Bigl({1 \over \delta+\rho^2} +
{1 \over \Delta+\rho^2} \Bigr)z_1\bar z_2,\xx && c_{22}={\rho^2
\over \Delta(\Delta+\rho^2)^2(\Delta+\rho^2+\zeta)}
\Bigl((\Delta+\rho^2)(z_1\bar z_1+\zeta) -
(\Delta+\rho^2+\zeta)z_2\bar z_2\Bigr),\xx &&
d_{11}=-{2(\rho^2+\zeta) \over \Delta+\rho^2} \sqrt{1\over
\delta\Delta(\Delta+\rho^2)(\Delta+\rho^2+\zeta)}\, \bar z_1\bar z_2,\\
&& d_{12}={2\over (\Delta+\rho^2)^2} \sqrt{\rho^2(\rho^2+\zeta)
\over \delta \Delta}\, \bar z_2 \bar z_2,\xx && d_{21}=-{2\over
\Delta+\rho^2+\zeta} \sqrt{\rho^2(\rho^2+\zeta) \over \Delta
(\Delta+\zeta)(\Delta+\rho^2)(\Delta+\rho^2+2\zeta)}\, \bar z_1
\bar z_1,\xx && d_{22}={2 \rho^2 \over \Delta+\rho^2+\zeta}
\sqrt{1 \over
\Delta(\Delta+\zeta)(\Delta+\rho^2)(\Delta+\rho^2+\zeta)}\, \bar
z_1 \bar z_2.\nonumber
\end{eqnarray}
One can check again the solution \eq{sdf2} is also well-defined on
the whole space $\do \times \re$ up to a mild singularity. The
above solution also reduces to the ordinary $SU(2)$ instanton in
the $\zeta = 0$ limit and the $U(1)$ solution \eq{sdf1} for the
limit $\rho =0$ where only $c_{11}$ and $d_{11}$ terms in
\eq{asdf2} survive.

One can check that the topological charge for the solution
\eq{sdf2} has exactly the same expression as \eq{asdQ2} except the
sign which is now $+$. So we can conclude for the same reason as
section 3.2 that the topological charge of the self-dual $U(2)$
instanton is also an integer and independent of the modulus
$\rho$. (Actually it should be since the changes of ADHM data in
\eq{ASD} and \eq{SD} are only $z_2 \leftrightarrow \bar z_2$ and
$z_1 \leftrightarrow -z_1$. However these changes should not be
important since $z_2$ and $\bar z_2$ are commutative coordinates,
i.e. $[\bar z_2, z_2]=0$.)

%%%%%%%%%%%%%%%%%%%%%%%%%%%%%%%%%%%%%%%%%%%%%%%%%%%%%%%%%%%%%%%%%%%%%%
\section{Discussion}
%%%%%%%%%%%%%%%%%%%%%%%%%%%%%%%%%%%%%%%%%%%%%%%%%%%%%%%%%%%%%%%%%%%%%%

In this letter we studied anti-self-dual and self-dual
noncommutative instantons on $\do \times \re$ based on the ADHM
construction. Unlike $\mi$, the ADHM equation for the
noncommuative space $\do \times \re$ is always deformed since this
system is not a BPS state any more. Remarkably, although the
instanton solutions for both self-dual and anti-self-dual gauge
fields contain a mild singularity, i.e. a measure zero
singularity, it always disappears in gauge invariant quantities
and thus physically regular solutions can be constructed even
though any projected states are not involved in the ADHM
construction. Furthermore the instanton number is always an
integer.

Our present result is different from \ct{ckt} by Chu, et al.
claiming that there is no nonsingular $U(N)$ instanton on ${\bf
R}^2_{NC} \times {\bf R}^2_{C}$ due to the breakdown of the
completeness relation. The authors of \ct{ckt} argued that if the
offending state, e.g. $|0\rangle$, is not subtracted, a huge plane
singularity on the whole $\do$-plane placed at $z_2=0$ is
developed in the solution and this huge singularity is not allowed
in the semi-classical picture, drawing the conclusion that the
vacuum structure of noncommutative $U(N)$ gauge theories on $\do
\times \re$ is trivial for all $N \geq 1$. However we showed that
although the solutions contain mild singularities, these
singularities always disappear whenever we define the gauge
invariant quantities, so they don't induce any physical
singularities. Also they appear only in the $U(1)$ part of $U(N)$
gauge theory. Thus the singularity in the $U(N)$ instanton
solution on ${\bf R}^2_{NC} \times {\bf R}^2_{C}$ is a gauge
artifact in the sense that it appears only gauge non-invariant
quantities.

The space ${\bf R}^2_{NC} \times {\bf R}^2_{C}$ can be realized
as the spatial worldvolume of D4-brane with rank-2 B field.
Obviously one can put D0-branes on this D4-brane. By SUSY
analysis, this system is not supersymmetric, so FI-term should be
introduced in the D4-brane world volume theory. This FI-term
appears as the deformation of ADHM equation as in \eq{ADHMr}. This
means that the D0-brane moduli space is resolved, i.e. the
D0-brane is a little bit smeared out on the D4-brane. This
picture is consistent with the result in this work. There is no
reason why such a huge plane singularity claimed in \ct{ckt}
should be developed in the D4-brane and why the D0-brane on the
D4-brane is so singular.

The instanton configurations on $\do\times \re$ can be naturally
explained by the topology of gauge group suggested by Harvey
\ct{harvey}, where the gauge transformations on $\do\times \re$
are characterized by the
maps from ${\bf S}^1$ to $U_{\mbox{cpt}}(\CH)$. Here
$U_{\mbox{cpt}}(\CH)$ denotes the unitary operators over $\CH$ of
the form $U={\bf 1}+K$ with $K$ a compact operator. This map is
nontrivial since $\pi_1(U_{\mbox{cpt}})={\bf Z}$. Thus the vacuum
structure of noncommutative gauge theory on $\do\times \re$ is
still parameterized by an integer winding number. This fact was
already noticed by Chu, {\it et al.} in \ct{ckt}, but rejected
for a wrong reason.

Multi-instanton solutions on $\do\times \re$ can be constructed
too. As shown in \ct{lty}, after separating out the center of
mass, the moduli space of two $U(1)$ instantons on $\mi$ is given
by the Eguchi-Hanson metric which is non-singular even at the
origin where the two $U(1)$ instantons coincide. As shown in this
paper the $U(1)$ instantons on $\do\times \re$ are definitely
non-singular. However, in our case, any projection is not
involved in the solution and the commutative space $\re$ is still
there. Thus it will be interesting to study whether or not these
differences can affect the moduli space for the two $U(1)$
instantons.

\section*{Acknowledgments}
We thank C.-S. Chu, V. V. Khoze and G. Travaglini for
helpful correspondence. HSY is grateful to the organizers, especially
Kimyeong Lee, of KIAS workshop on ``Solitonic Objects in String and
Field Theories'' (September 10-14, 2001, Seoul, Korea) for providing
him an opportunity
to present this work and the support during his visit.
KYK and BHL are supported by the Ministry
of Education, BK21 Project No. D-0055 and by grant No.
1999-2-112-001-5 from the Interdisciplinary Research Program of
the KOSEF. HSY is supported by NSC (NSC90-2811-M-002-019). He also
acknowledge NCTS as well as CTP at NTU for partial support.

\newpage

%%%%%%%%%%%%%%%%% Journal Macros %%%%%%%%%%%%%%%%%%%%%%%%%%%

\nc{\np}[3]{Nucl. Phys. {\bf B#1} (#2) #3}

\nc{\pl}[3]{Phys. Lett. {\bf B#1} (#2) #3}

\nc{\prl}[3]{Phys. Rev. Lett. {\bf #1} (#2) #3}

\nc{\prd}[3]{Phys. Rev. {\bf D#1} (#2) #3}

\nc{\ap}[3]{Ann. Phys. {\bf #1} (#2) #3}

\nc{\prep}[3]{Phys. Rep. {\bf #1} (#2) #3}

\nc{\ptp}[3]{Prog. Theor. Phys. {\bf #1} (#2) #3}

\nc{\rmp}[3]{Rev. Mod. Phys. {\bf #1} (#2) #3}

\nc{\cmp}[3]{Comm. Math. Phys. {\bf #1} (#2) #3}

\nc{\mpl}[3]{Mod. Phys. Lett. {\bf #1} (#2) #3}

\nc{\cqg}[3]{Class. Quant. Grav. {\bf #1} (#2) #3}

\nc{\jhep}[3]{J. High Energy Phys. {\bf #1} (#2) #3}

\nc{\hep}[1]{{\tt hep-th/{#1}}}

%%%%%%%%%%%%%%%%%%%%%%%%%%%%%%%%%%%%%%%%%%%%%%%%%%%%%%%%%%%


\begin{thebibliography}{99}


\bibitem{adhm} M. Atiyah, N. Hitchin, V. Drinfeld and Y. Manin,
Phys. Lett. {\bf 65A} (1978) 185; E. Corrigan and P. Goddard, \ap
{154}{1984}{253}; S. Donaldson, \cmp {93}{1984}{453}.


\bibitem{ns} N. A. Nekrasov and A. Schwarz, \cmp {198}{1998}{689},
\hep{9802068}.


\bibitem{mb} M. Berkooz, \pl {430}{1998}{237}, \hep{9802069}.


\bibitem{naka} H. Nakajima, Ann. of Math. {\bf 145} (1997) 379;
Nucl. Phys. Proc. Suppl. {\bf 46} (1996) 154.


\bibitem{sw} N. Seiberg and E. Witten, \jhep {09}{1999}{032}, \hep{9908142}.


\bibitem{ly} K. Lee and P. Yi, \prd {61}{2000}{125015},
\hep{9911186}.


\bibitem{bn} H. W. Braden and N. A. Nekrasov, Spacetime Foam
from Noncommutative Instantons, \hep{9912019}.


\bibitem{kf1} K. Furuuchi, \ptp{103}{2000}{1043}, \hep{9912047}.


\bibitem{kf2} K. Furuuchi, \cmp {217}{2001}{579}, \hep{0005199};
Topological Charge of U(1) Instantons, \hep{0010006}.


\bibitem{kf3} K. Furuuchi, \jhep {03}{2001}{033}, \hep{0010119}.


\bibitem{kko} A. Kapustin, A. Kuznetsov and D. Orlov,
\cmp {221}{2001}{385}, \hep{0002193}.

\bibitem{ho} P.-M. Ho, Twisted Bundle on Noncommuative Space and
$U(1)$ Instanton, \hep{0003012}.


\bibitem{kly} K.-Y. Kim, B.-H. Lee and H. S. Yang, Comments on Instantons
on Noncommutative ${\bf R}^4$, \hep{0003093}.


\bibitem{lty} K. Lee, D. Tong and S. Yi, \prd {63}{2001}{065017},
\hep{0008092}.


\bibitem{nek} N. A. Nekrasov,
Trieste lectures on solitons in noncommutative gauge theories, \hep{0011095}.


\bibitem{as} A. Schwarz, \cmp {221}{2001}{433}, \hep{0102182}.


\bibitem{corr} D. H. Correa, G. S. Lozano, E. F. Moreno
and F. A. Schaposnik, Comments on the U(2) Noncommutative
Instanton, \hep{0105085}.


\bibitem{ckt} C.-S. Chu, V. V. Khoze and G. Travaglini,
Notes on Noncommutative Instantons, \hep{0108007}.


\bibitem{ham} M. Hamanaka, ADHMN/Nahm Construction of Localized
Solitons in Noncommutative Gauge Theories, \hep{0109070}.


\bibitem{kuro} T. Ishikawa, S.-I. Kuroki and A. Sako,
Elongated U(1) Instantons on Noncommutative ${\bf R^4}$,
\hep{0109111}.


\bibitem{harvey} J. A. Harvey, Topology of the Gauge Group in
Noncommutative Gauge Theory, \hep{0105242}.

\newpage
%%%%%%%%%%%%%%%%%%%%%%%%%%%%%%%%%%%%%%%%%%%%%%%%%%%%%
\begin{figure}[tbp]
\centerline{\epsfxsize=18cm \epsfbox{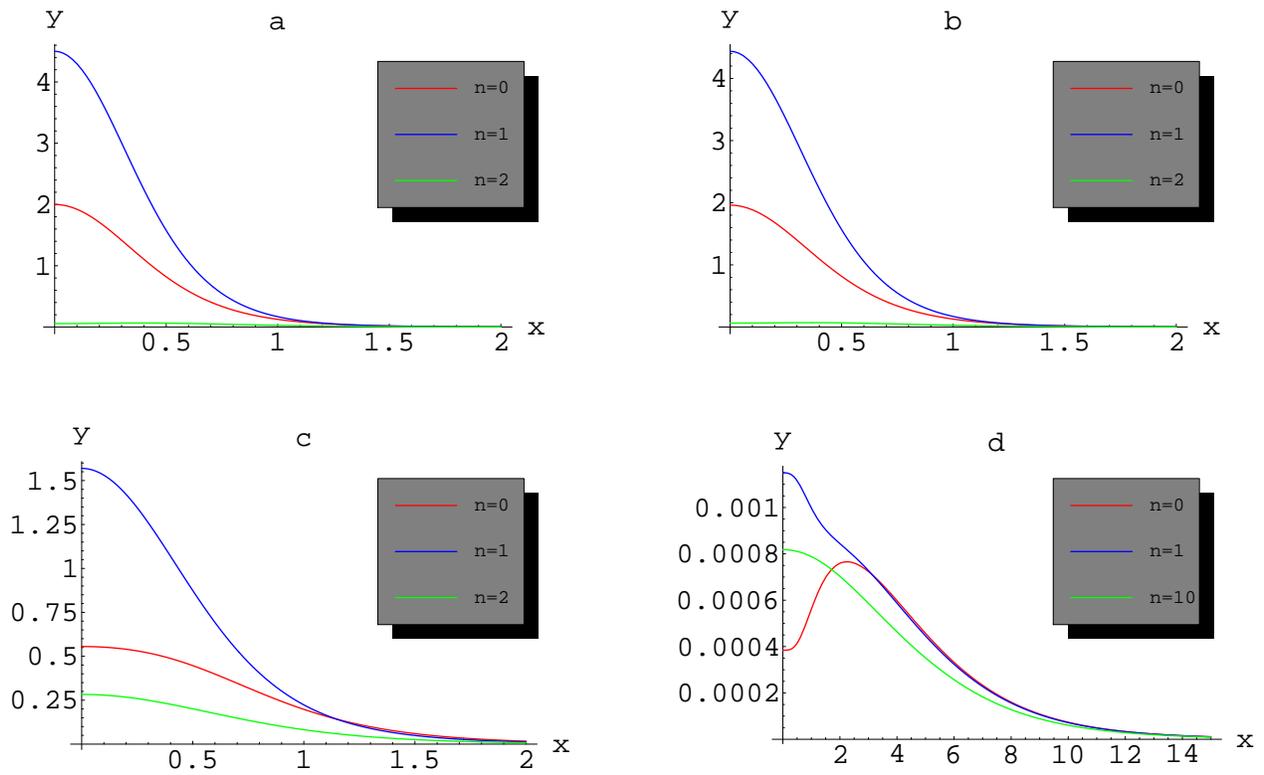}} \caption{
Topological charge densities $y=2Q_n(\gamma)$
  defined in \eq{asdQ1}
  and \eq{densityQ} where $x=\sqrt{\gamma}$. a: $U(1)$, b: $U(2),\;a=0.1$,
c: $U(2),\;a=1$,  d: $U(2),\;a=10$.} \label{shape}
\end{figure}
%%%%%%%%%%%%%%%%%%%%%%%%%%%%%%%%%%%%%%%%%%%%%%%%%%%%%



\end{thebibliography}
\end{document}